\documentclass[10pt,conference]{IEEEtran}
\usepackage{tikz-cd}
\usetikzlibrary{arrows}
\usepackage{scalerel}
\newcommand{\rimp}{\Rightarrow}
\usepackage{amsfonts} 
\usepackage{url}
\usepackage{algpseudocode}
\algnewcommand\algorithmicforeach{\textbf{for each}}
\algdef{S}[FOR]{ForEach}[1]{\algorithmicforeach\ #1\ \algorithmicdo}

\usepackage{lipsum}

 \newtheorem{definition}{Definition}
 \newtheorem{problem}{Problem}
 \newtheorem{example}{Example}
  \newtheorem{theorem}{Theorem}

\usepackage{float}
\floatstyle{boxed}
\newfloat{algorithm}{htbp}{loa}
\floatname{algorithm}{Algorithm}
\newfloat{algorithm}{htbp}{loa}
\usepackage{wrapfig}
\usepackage{multirow}
\usepackage{amsmath,epsfig,epstopdf,amssymb}

\usepackage{enumerate}
\usepackage{setspace}

\usepackage{lipsum}
\usepackage{mathtools}
\usepackage{cuted}

\usepackage[colorinlistoftodos,prependcaption,textsize=small]{todonotes}
\usepackage{hyperref}
\usepackage{tikz}
\usetikzlibrary{matrix}

\title{Efficient Decentralized LTL Monitoring Framework Using Tableau Technique}

\author{Omar Al-Bataineh and David Rosenblum \\ School of Computing \\
 National University of Singapore}

\date{}

\begin{document}

\maketitle

\begin{abstract}

This paper presents a novel framework for 
decentralized monitoring of  Linear   Temporal Logic (LTL), 
under the situation where processes are synchronous
and the formula is represented as a tableau.
The tableau technique allows one to construct
a semantic tree for the input formula, which can be used  
to optimize the decentralized monitoring of LTL  in various ways.
Given a system $P$ and an LTL formula $\varphi$, 
we construct a tableau $\mathcal{T}_{\varphi}$.
The tableau $\mathcal{T}_{\varphi}$ is used for two purposes:
(a) to synthesize an efficient round-robin communication policy
for processes, and (b) to find the minimal ways to decompose the global formula
and communicate partial observations in an optima way.
In our framework, processes can propagate truth values of atomic formulas,  compound formulas, and temporal formulas depending on the syntactic structure of the input LTL formula and the observation power of processes.  
We demonstrate that this approach of decentralized monitoring  based on tableau construction is more straightforward, more flexible, and more likely to yield efficient solutions than alternative approaches.

\end{abstract}

\section{Introduction}

Run-time verification (RV) has been recognized as one of the integral parts of software and hardware design process. RV is a lightweight formal method that aims to verify (at runtime) the conformance of the executions of the system under analysis with respect to some desired properties. Typically the system is considered as a black box that feeds events to a monitor. An event usually consists of a set of atomic propositions that describe some abstract operations in the system.  Based on the received events, 
the monitor emits verdicts in a truth domain that indicate whether or not the run
complies with the specification. The technique has been successfully applied to a number of industrial software systems, providing extra assurance of behavior correctness \cite{ColomboPS08,Pike2011,DAngeloSSRFSMM05}.

 Recently, there has been a growing interest in run-time verification of distributed systems, that is, systems with multiple processes and no central observation
point. Building a decentralized runtime monitor for a distributed system is a non-trivial task since it involves designing a distributed algorithm that coordinates the monitors in order to reason consistently about the temporal behavior of the system.  
The key challenge is that the monitors have a partial
view of the system and need to account for communication and consensus.

In this work, we consider the decentralized monitoring of systems
under the following setting: (a)  processes are synchronous, 
and (b) the formula is expressed in a high-level 
 requirement  specification written  in LTL and represented as a tableau.
The tableau technique \cite{Reynolds2016} has  many  applications  in  logic.  It  is  used  as  a  method  of  verifying  whether  a  given  formula  is  a  tautology,  as  a  method  of  proving  semantical  consistency  of  a  set  of  formulas,  and  even  as  an  algorithm  for  verifying of the validity of arguments.  
 Using the tableau technique in decentralized monitoring has several advantages. 
 First,  by constructing a tableau for the monitored formula, we can detect early 
 whether the formula is a tautological formula or unsatisfiable formula with zero communication overhead. Second, it allows processes to propagate information  about only \textit{feasible} branches of the tableau (i.e. successful branches), where no information will be propagated about \textit{infeasible} branches (i.e. failed branches). 
Third, it helps to find the minimal way to decompose the formula
and communicate partial observations of processes in an efficient way.

The presented decentralized framework 
 consists mainly of two parts: (a) a tableau-based 
algorithm that allows processes to compute at each state of the input execution trace the minimal set of formulas whose truth values need to be propagated, and (b) an LTL inference engine that allows processes to extract the maximal amount of information from a received message in case  the message contains truth values of non-atomic formulas.
We demonstrate that this approach of decentralized monitoring  based on tableau construction is more straightforward, more flexible, and more likely to yield efficient solutions than alternative approaches.

\paragraph{\textbf{Contributions}} We summarize our contributions
in this work as follows.

\begin{itemize}

\item We present a new decentralized monitoring framework
for LTL formulas under the assumption where processes
are synchronous and the formula is represented as a tableau. 
The framework inherits known advantages of the tableau technique,
LTL inference rules, and static monitoring approaches based on round-robin policies.

\item 
We show how the tableau technique can be used to optimize 
decentralized monitoring of LTL
(a) by finding the minimal ways to represent and decompose the global formula,
and (b) by synthesizing efficient round-robin communication policies based on the semantics of the constructed tableau of the monitored formula.

\item We develop an LTL inference engine that takes into consideration
the observation power of processes and the syntactic structure of the compound formula being analyzed.
The engine is used by processes to extract the maximal allowed of information
from the received messages.
The inference engine operates with three categories of rules:
(a) inference rules for propositional logic that
 are not dependent on the observation power of processes, 
 (b) inference rules for propositional logic that are 
 dependent on the observation power of processes, 
 and (c) inference rules for temporal logic.

\end{itemize}

\section{Preliminaries}

 In this section, we give a brief review of the syntax and semantics
of linear temporal logic LTL and the decentralized LTL monitoring problem.
Then we review the basic tableau decomposition/expansion rules for LTL.

\subsection{Decentralized LTL Monitoring Problem}

A distributed program $\mathcal{P} = \{p_1,p_2,...,p_n \}$ is a set of $n$ processes which cooperate with each other in order to achieve a certain task. 
Distributed monitoring is less developed and more challenging than local monitoring: they involve designing a distributed algorithm that monitors another distributed algorithm.
In this work,  we assume that no two processes share a common variable.  Each process of the distributed system emits events at discrete time instances. Each event $\sigma$ is a set of actions denoted by some atomic propositions from the set $AP$. We denote $2^{AP}$ by $\Sigma$ and call it the alphabet of the system. We assume that the distributed system operates under the perfect synchrony hypothesis, and that each process sends and receives messages at \textit{discrete} instances of time, which are represented using identifier $t \in \mathbb{N}^{\geq 0}$. An event in a process $p_i$, where $1 \leq i \leq n$, is either

\begin{itemize}

\item  internal event (i.e. an assignment statement),

\item  message sent, where the local state of $p_i$ remains unchanged, or

\item  message received,  where the local state of $p_i$ remains unchanged. 

\end{itemize}

Since each process sees only a projection of an event to its locally observable set of actions, we use a projection function $\Pi_i$ to restrict atomic propositions to the local view of monitor $\mathcal{M}_i$ attached to process $p_i$, which can only observe those of process $p_i$. For atomic propositions (local to process $p_i$), $\Pi_i: 2^{AP} \rightarrow 2^{AP}$, and we denote $AP_i = \Pi_i (AP)$, for all $i =1...n$. For events, $\Pi_i :2^{\Sigma} \rightarrow 2^{\Sigma}$ and we denote $\Sigma_i = \Pi_i (\Sigma)$ for all $i= 1...n$. We assume that $\forall_{i, j \leq n, i \neq j}  \rimp AP_i \cap AP_j = \emptyset$ and consequently  $\forall_{i, j \leq n, i \neq j} \rimp \Sigma_i \cap \Sigma_j = \emptyset$. That is, events are local to the processes where they are monitored.
The system's global trace, $g = (g_1, g_2,..., g_n)$ can now be described as a sequence of pair-wise unions of the local events of each process's traces. We denote the set of all possible events in $p_i$ by $E_i$ and hence the set of all events of $P$ by $E_P = \bigcup_{i=1}^{n} E_i$. Finite traces over an alphabet $\Sigma$ are denoted by $\Sigma^{*}$, while infinite traces are denoted by $\Sigma^{\infty}$.

\begin{definition}(LTL formulas \cite{Pnueli1977}). The set of LTL formulas is inductively defined by the grammar
\[
\varphi ::=  true \mid p \mid \neg \varphi \mid \varphi \lor \varphi \mid X \varphi \mid F \varphi \mid G \varphi \mid \varphi U \varphi  
\]
where $X$ is read as next, $F$ as  eventually (in the future), 
$G$ as  always (globally), $U$ as until, and $p$ is a propositional variable.

\end{definition}

\begin{definition} (LTL Semantics \cite{Pnueli1977}). Let $ w = a_0 a_1.. . \in \Sigma^{\infty}$ be a infinite word with $i \in N$
being a position. Then we define the semantics of LTL formulas inductively as follows

\begin{itemize}

\item $w, i \models true$

\item $w, i \models \neg \varphi$ iff $w, i \not\models \varphi$

\item $w, i \models p$ iff $ p \in a_i$
 
\item  $w, i \models \varphi_1 \lor \varphi_2$ iff $w, i \models \varphi_1$ or $w, i \models \varphi_2$

\item $w, i \models F \varphi$ iff $ w, j \models \varphi$ for some $j \geq i$

\item $w, i \models G \varphi$ iff $ w, j \models \varphi$ for all $j \geq i$

\item $w, i \models \varphi_1 U \varphi_2$ iff $\exists_{k \geq i}$ with $w, k \models \varphi_2$ and $\forall_{i \leq l < k}$ with $ w, l \models \varphi_1$

\item  $w, i \models X \varphi$ iff $w, i+1 \models  \varphi$

\end{itemize}

\end{definition}

We now review the definition of three-valued semantics LTL$_3$ that is used to interpret common LTL formulas,
as defined in \cite{Bauer2011}. The semantics of  LTL$_3$ is defined on finite prefixes to obtain a truth value from the set $\mathbb{B}_3 = \{ \top, \bot, ?  \}$.

\begin{definition} (LTL$_3$ semantics). Let $u \in \Sigma^{*}$ denote a finite word.  The truth value of a LTL$_3$ formula $\varphi$ with respect to $u$, denoted by $[u \models \varphi]$, is an element of $\mathbb{B}_3$ defined as follows:

$$
[u \models \varphi]  = 
\begin{cases}
\top & \textrm{if $\forall \sigma\in \Sigma^{\infty} : u\sigma \models \varphi $} \\
 \bot & \textrm{if  $\forall \sigma\in \Sigma^{\infty} : u\sigma \not\models \varphi $} \\
 ? & otherwise
\end{cases}
$$

\end{definition}

Note that according to the semantics of LTL$_3$ the outcome of the evaluation of $\varphi$ can be inconclusive (?). This happens if the so far observed prefix $u$ itself is insufficient to determine how $\varphi$ evaluates in any possible future continuation of $u$.

\begin{problem} (The decentralised monitoring problem).
Given a distributed program $\mathcal{P} = \{p_1, p_2,..., p_n\}$, a finite global-state trace $\alpha \in \Sigma^{*}$, an $LTL$ property $\varphi$, and a set of monitor processes $ \mathcal{M} = \{ M_1, M_2,..., M_n\}$ such that

\begin{itemize}

\item  monitor $M_i$ can read the local state of process $p_i$, and

\item  monitor $M_i$ can communicate with other monitor processes.


\end{itemize}
The problem is then to design an algorithm that allows each monitor $M_i$ to evaluate $\varphi$ through communicating with other monitor processes.
The problem can be studied under different settings and different assumptions.
However, in this work,
we make a number of assumptions
about the class of systems that can be monitored in our framework.

\end{problem}

\begin{itemize}

\item \textbf{A1}: the monitored system  is a synchronous timed system with a global clock;


\item \textbf{A2}: processes are reliable (i.e., no process is malicious).
\end{itemize}

It is interesting to note that the synchronous assumption imposed in our setting
is by no means unrealistic, as in many real-world systems,
communication occurs synchronously. We refer the reader to \cite{BauerF12,ColomboF14} 
in which the authors
discussed a number of interesting examples of protocols 
for safety-critical systems in which communication occurs synchronously. Examples include the FlexRay bus protocol \cite{Pop06timinganalysis,Gunzert1999} and Deterministic Ethernet (cf. IEEE802.1 or \cite{Oliver2014AnalysisOD} for an overview).

\subsection{Tableau Construction for LTL} \label{sec:tableau}

Among the various existing tableau systems for LTL, we selected Reynolds's implicit
declarative one \cite{Reynolds2016}. 
The tableau is unique in that it is wholly traditional in style (labels are sets of formulas),
it is tree shaped tableau construction, 
and it can handle repetitive loops and infinite branches quite efficiently.

Given an LTL formula $\varphi$ we construct a direct graph (tableau) $\mathcal{T}_{\varphi}$ using the standard expansion rules for LTL. 
Applying expansion rules to a formula leads to a new formula but with an equivalent semantics. 
We review here the basic expansion rules of temporal logic:  (1) $G p \equiv p \land XG p$,
(2) $F p \equiv p \lor XF p$, and (3) $p ~ U q \equiv q \lor (p \land X(p~ U q)).$ 
Tableau expansion rules for propositional logic are very straightforward 
and can be described as follows:

\begin{itemize}

\item If a branch of the tableau contains a conjunctive formula $A \land B$, 
add to its leaf the chain of two nodes containing the formulas $A$ and $B$.

\item   If a node on a branch contains a disjunctive formula $A \lor B$, 
then create two sibling children to the leaf of the branch, 
containing $A$ and $B$, respectively.

\end{itemize}

The aim of tableaux is to generate progressively simpler
formulas until pairs of opposite literals are produced
or no other expansion rule can be applied.
The labels on the tableau proposed by Reynolds are just sets of
formulas from the closure set of the original
formula. Note that one can use De Morgan's laws during
the expansion of the tableau, so that for example,
 $\neg (a \land b)$ is treated
as $\neg a \lor \neg b$.
A node in $\mathcal{T}_{\varphi}$ is called a leaf if it  has zero-children. 
A leaf may be crossed (x), indicating its branch has failed 
(i.e., contains opposite literals), or ticked $\surd$, 
indicating its branch is successful. The whole tableau $\mathcal{T}_{\varphi}$ 
is successful if there is a successful branch, and in this case we
say that the formula  $\varphi$ is satisfiable in the sense that there exists a model
that satisfies the formula.

\begin{figure} [h]
   \begin{minipage}[b]{0.45\linewidth}
    \begin{center}
    \begin{tikzcd}  
 & p \land (q \lor r) \arrow{d}  \\
	 & p, (q \lor r) \arrow{ld} \arrow{rd} \\
    p, q   & & p, r  \\
    \surd  & & \surd
 \end{tikzcd}  
     \end{center}
     \caption{A tableau for  $(p \land (q \lor r))$}
        \end{minipage}
   \begin{minipage}[b]{0.7\linewidth}
    \begin{center}
    \begin{tikzcd}  
   Gp \arrow{d}   \\
   p, XGp \arrow{d}   \\
   Gp \arrow{d}  \\
      p, XGp   \\
      \surd
 \end{tikzcd}  
     \end{center}
     \caption{A tableau for $Gp$} \label{Fig:EGP}
   \end{minipage}
   \end{figure}

Note that Reynolds \cite{Reynolds2016} introduced
a number of new  tableau rules which support a new simple traditional
style tree-shaped tableau for LTL. 
The new rules (the PRUNE rule and the LOOP rule)
provide a simple way to curtail repetitive branch
extension. We therefore guarantee that the tableau construction always terminates
and returns a semantic graph for the monitored formula
including those containing nested operators.
For example, the formula $G p$ (see Fig. \ref{Fig:EGP})
gives rise to a very repetitive infinite tableau without the LOOP rule, 
but succeeds quickly with it. Note that the label $(p, XGp)$
 is repeated two times and hence a fixed-point is reached.
In this case we stop the analysis and declare that the tableau is successful.

\section{Decentralized LTL Inference Engine} \label{sec: inferenceRules}

In this section we describe an inference engine (a set of IF-THEN inference rules)  
which consists of a number of inference rules on propositional logic
that can be used by processes to deduce definite truth
values of propositions in compound formulas. 
The engine performs syntactic decomposition 
of the compound formula using tableau decomposition rules. 
Before formalizing the inference rules, we introduce some notations:

\begin{itemize}

\item $AP_p$, the set of locally observed proposition by process $p$,
and $obs(p)$ be the set of propositions 
whose definite truth values are known to process $p$ either by direct
observation or remote observation through communication;

\item $LOP (\phi)$, the set of logical operators of formula $\phi$;

\item$TOP (\phi)$, the set of temporal operators of formula $\phi$; and

\item $Atoms(\phi)$, the set of atomic propositions in $\phi$.

\end{itemize}

Our decentralized monitoring algorithm is mainly based on a three-valued semantics for the future linear temporal logic LTL, where the interpretation of the third truth value, denoted by ?, follows Kleene logic \cite{Kleene29}.  For example, a monitor might not know the Boolean value of a proposition at a time point because it is not within its observation power. In this case, the monitor assigns the proposition the truth value ?. Note that no verdict is produced if under the current knowledge the specification evaluates to ?.  We can then summarize the basic cases for three-valued logical operators involving (?) as follows: ($\top  \land ~? = ? $),($\bot  \land ~? = \bot $), ($?~  \land ~ ? = ? $), ($\top  \lor ~? = \top $), ($\bot  \lor ~? = ? $), ($?~  \lor ~ ? = ? $), ($\neg ~ ?  = ?$). 
Rules of inference are written in the following form 
$$
\begin{tikzpicture}
  \draw (-2,0) -- node[below] {$Conclusion_1 \land  ... \land Conclusion_n$} node[above] {$Premise_1;...; Premise_n$} ++(5,0);
\end{tikzpicture}
$$
where $Premise_1;..., Premise_n$ are a list of premises 
and $Conclusion_1; ...; Conclusion_n$ are the list of logical consequences  that can be derived from  $Premise_1;..., Premise_n$. We write $(\phi)^{v}_{p}$
to denote that the formula $\phi$ has been evaluated to $v \in \mathbb{B}_3$ by process $p$.
We classify the inference rules into three categories: (a) inference rules
on propositional logic that are not dependent on the observation power
of processes, (b) inference rules
on propositional logic that are dependent on the observation power, 
and (c) inference rules on temporal logic.

\begin{itemize}

\item  \textbf{Inference rules
on propositional logic not dependent on the observation power
of processes}. Rules  \textbf{R1}-\textbf{R2} are
 straightforward rules which take advantage of the semantics
of logical operators $(\land, \lor)$ to deduce definite 
truth values of propositions in a formula.
$$  \textbf{R1}:
\begin{tikzpicture}
  \draw (-2,0) -- node[below] {$ (\forall_{i = 1...n } (a_i = \top))$} node[above] {$(a_1 \land a_2 \land ... \land a_n)^{\top}$} ++(3,0);
\end{tikzpicture}
 \textbf{R2}:
\begin{tikzpicture}
  \draw (-2,0) -- node[below] {$(\forall_{i =1 ...n } (a_i = \bot))$} node[above] {$(a_1 \lor a_2 \lor ... \lor  a_n)^{\bot}$} ++(3,0);
\end{tikzpicture}
$$

\item \textbf{Inference rules on propositional logic dependent on the observation power
of processes}.  An interesting set of inference rules on propositional logic 
can be developed if we allow receiving processes to take into consideration the observation power of the sending processes when extracting information from the received compound formulas,  as described in rules \textbf{R3-R10}.  
The rules given at this category are straightforward rules except rules  \textbf{R9} and  \textbf{R10}, which we will explain later by an example. The rules work as follows: 
depending on the syntactic structure of the compound formula $\phi$ 
and its truth value as  evaluated by a sending process $p$,
a receiving process $q$ can deduce definite truth values of some propositions in $\phi$.
Therefore, these rules should be viewed as inference rules used by process $q$ after receiving a message from process $p$ containing truth values of compound formulas.
Recall that we assume here that processes know the observation power of each other. 
$$
 \textbf{R3}:
\begin{tikzpicture}
  \draw (-2,0) -- node[below] {$(\forall_{a_i \in obs(p)} (a_i = \top))_{q}$} node[above] {$(a_1 \land a_2 \land ... \land a_n)^{?}_{p};~ (\{a_1,..., a_n\} \cap obs(p)) \neq \emptyset  $} ++(7,0);
\end{tikzpicture}
$$
$$
 \textbf{R4}:
\begin{tikzpicture}
  \draw (-2,0) -- node[below] {$(\forall_{a_i \in obs(p)} (a_i = \bot))_{q}$} node[above] {$(a_1 \lor a_2 \lor ... \lor a_n)^{?}_{p};~ (\{a_1,..., a_n\} \cap obs(p)) \neq \emptyset  $} ++(7,0);
\end{tikzpicture}
$$
$$
\resizebox{\linewidth}{!} {
\textbf{R5}:
\begin{tikzpicture}
  \draw (-2,0) -- node[below] {$(a = \bot \land b = \bot)_{q}$} node[above] {$(a \lor (b \land c))^{\bot}_{p};~ a, b \in obs(p)  $} ++(4,0);
\end{tikzpicture}
\textbf{R6}:
\begin{tikzpicture}
  \draw (-2,0) -- node[below] {$(a = \bot \land b = \top)_{q}$} node[above] {$(a \lor (b \land c))^{?}_{p};~ a, b \in obs(p)  $} ++(4,0);
\end{tikzpicture}}
$$

$$
\resizebox{\linewidth}{!} {
 \textbf{R7}:
\begin{tikzpicture}
  \draw (-2,0) -- node[below] {$(a = \top \land b = \bot)_{q}$} node[above] {$(a \land (b \lor c))^{?}_{p};~ a, b \in obs(p)  $} ++(4,0);
\end{tikzpicture}
\textbf{R8}:
\begin{tikzpicture}
  \draw (-2,0) -- node[below] {$(a = \top \land b = \top)_{q}$} node[above] {$(a \land (b \lor c))^{\top}_{p};~ a, b \in obs(p)  $} ++(4,0);
\end{tikzpicture}}
$$

$$
\resizebox{\linewidth}{!} {
 \textbf{R9}:
\begin{tikzpicture}
  \draw (-2,0) -- node[below] {$  \begin{array}[t]{l}(\forall_{i =1..n} (\forall_{a \in (Atoms(\phi_i) \cap obs(p))} (a = \bot)) \land  \\ \forall_{i =1..n} (\forall_{\neg a \in (\overline{Atoms}(\phi_i) \cap obs(p))}) (a = \top)))_{q} \end{array}$} node[above] {$ \begin{array}[t]{l}
(\phi_1 \lor ...\lor \phi_n)^{\bot}_{p}; \forall_{i =1..n} (TOP(\phi_i) = \emptyset); \forall_{i =1..n} (LOP(\phi_i) = \{\land\}; \\  ~\forall_{i =1..n} (Atoms(\phi_i) > 1); \forall_{i =1..n} ((|Atoms(\phi_i) \cap obs(p)| = 1) ) \end{array}
$} ++(10,0);
\end{tikzpicture}}
$$

$$
\resizebox{\linewidth}{!} {
 \textbf{R10}:
\begin{tikzpicture}
  \draw (-2,0) -- node[below] {$\begin{array}[t]{l} (\forall_{i =1..n} (\forall_{a \in (Atoms(\phi_i) \cap obs(p))} (a = \top)) \land \\ \forall_{i =1..n} (\forall_{\neg a \in (\overline{Atoms}(\phi_i) \cap  obs(p))}) (a = \bot)))_{q} \end{array} $} node[above] {$ \begin{array}[t]{l}
(\phi_1 \land ...\land \phi_n)^{\top}_{p}; \forall_{i =1..n} (TOP(\phi_i) = \emptyset); \forall_{i =1..n} (LOP(\phi_i) = \{\lor\}; \\  ~\forall_{i =1..n} (Atoms(\phi_i) > 1); \forall_{i =1..n} ((|Atoms(\phi_i) \cap obs(p)| = 1) ) \end{array}
$} ++(10,0);
\end{tikzpicture}
}
$$

We now turn to discuss rules \textbf{R9} and \textbf{R10}. Note that rule \textbf{R10} is the dual of
rule \textbf{R9} and we therefore explain only \textbf{R9}. 
The rule can be used if the syntactic structure of the complex compound formula
$\Phi = (\phi_1 \lor ...\lor \phi_n)$ satisfies the following conditions:
(a) the only logical operator appearing in subformula $\phi_i$ is $\{\land\}$,
(b) $\phi_i$ has no temporal operator
(c) process $p$ observes at most one proposition in each subformula $\phi_i$,
and (d) $\Phi$ is evaluated to $\bot$ by $p$. Then once a process $q$ receives
a message from $p$ containing the truth value of $\Phi$ it can deduce
that all positive propositions in $\Phi$ observed by $p$ have truth values $\bot$
and all negative propositions in $\Phi$ observed by $p$ have truth values $\top$.
For example, consider $\Phi = ( (a \land b) \lor (\neg c \land d) )$, 
where process $p$ observes $\{a, c \}$ and evaluates $\Phi$ to $\bot$.
Then once $q$ receives a message from $p$ containing the truth value of $\Phi$ 
it can  deduce that $a = \bot$ and $c = \top$.

\item \textbf{Inference rules for temporal logic}. 
We discuss here two inference rules for temporal operators.
Note that to develop useful inference rules for temporal logic
 we choose to restrict the scope of temporal operators to 
a specific time step $n$, where $n$ can be chosen to be the current time
step at which the formula is evaluated.
Such restriction allows us to reduce temporal formulas
into propositional formulas by using unfolding rules
for temporal operators.
For example, we can unfold the formula $G^{(n)} (\phi)$ to 
 $(\phi^{(0)} \land ... \land \phi^{(n)})$ 
given the semantics of the  operator $G$,
so that we apply the unfolding rule of the $G$ operator  $n$-times.
Similar unfolding rules can be applied 
to formula $F^{(n)} (\phi)$.
The resulting unfolding formulas are compound propositional formulas. 

$$
 \textbf{R11}: \begin{tikzpicture}
  \draw (-2,0) -- node[below] {$ (\forall_{t =0 ... n} (\phi^{(t)} = \top)) $} node[above] {$  (G^{(n)} (\phi))^{\top}$} ++(3,0);
\end{tikzpicture}
~~\textbf{R12}:
\begin{tikzpicture}
  \draw (-2,0) -- node[below] {$(\forall_{t =0... n} (\phi^{(t)} = \bot))$} node[above] {$  (F^{(n)} (\phi))^{\bot}$} ++(3,0);
\end{tikzpicture} 
$$

\end{itemize}

\noindent The advantage of having inference rules on temporal logic
as given in rules \textbf{R11-R12} is  that they allow distributed processes to deduce
information (i.e. truth values of propositions in the global LTL formula)
not just in one particular  state of the system, 
but also in a sequence of states. That is,
from a truth value of a single temporal formula, processes may
deduce information about $n$ past states.
This is extremely useful in decentralized monitoring,
as it allows processes to propagate their observations
in a more efficient way. 
To demonstrate the usefulness of these rules
in decentralized monitoring, consider  the following example.

\begin{example} \label{TempInfereRules}

Suppose we have a temporal formula $\Phi =  F (a \lor b \lor c)$
which is monitored in a decentralized fashion
by a group of processes $\{A, B, C\}$.
Let us denote $(a \lor b \lor c)$ by $\phi$.
Suppose that at time step $t > 0$ 
process $A$ told process $B$ that $\Phi^{(t)} = \bot$. 
Then from the syntactic structure of $\Phi$ and the semantics
of the temporal operator $F$ (see rule \textbf{R12}), process
$B$ can deduce that $\phi^{(0)} = \bot \land ... \land \phi^{(t)} = \bot$.
Since $\phi$ is a compound formula whose syntactic structure
matches the one given in rule \textbf{R2}, $B$ needs to apply
 \textbf{R2} on $\phi$ from $\tau = 0$ to $\tau = t$.
It then deduces that the truth values of the atoms $a, b$ and 
$c$ from $\tau = 0$ to $\tau = t$ are $\bot$. 
That is, $a^{(\tau)} = b^{(\tau)} = c^{(\tau)} = \bot$ for all $\tau \in [0, t]$.

\end{example}

In some cases processes may need to apply 
multiple inference rules on the same compound
formula depending on the syntactic structure of the formula. 
So that if the compound formula can be decomposed into multiple
compound formulas of the forms given in rules \textbf{R1}-\textbf{R10},
then multiple inference rules will be applied to derive
definite truth values of the atomic propositions in the original compound formula.
Consider the following example to demonstrate this. 

\begin{example} \label{ComplexCompoundFormula}

Suppose we have a compound formula $\Phi = ( (a \land b) \lor (c \land (d \lor e)))$
and that a process $p$ has $AP_p = \{a, d, e\}$.
Suppose that at time step $t$ the process $p$ observes that $a = d = e = \bot$ 
and hence evaluates the formula $\Phi$ to $\bot$.
Suppose further that $p$ propagates the truth value
of  $\Phi$  to some other process $q$ which knows that $p$ observes the atoms  $\{a, d, e\}$.
Note that $\Phi$ can be decomposed into two
compound subformulas $\phi_1 = (a \land b)$ and $\phi_2 = (c \land (d \lor e))$.
Since $p$ told $q$ that $\Phi = \bot$ then $q$ can deduce 
that $\phi_1 = \bot$ and $\phi_2 = \bot$. 
In this case process $q$ can deduce that $a = \bot \land  d = \bot \land e = \bot$.

\end{example}

\section{Computing Observation Power of Processes} \label{sec:ComutingObsPower}

In this section, we discuss the problem of computing the observation power
of distributed processes in a decentralized fashion, the set $obs^{(t)}(p_i)$ 
(i.e. the set of atomic propositions whose truth value are known to process $p_i$ at time step $t$).
The first intuitive way of computing the observation power of a process $p_i$ 
is to assume that $p_i$ knows the truth values of
 its local atomic propositions from the initial time step up to the current time step.
 We formalize this as follows

$$
\resizebox{\linewidth}{!} {
 \begin{array}[t]{l}
obs^{(t)}(p_i) = obs^{(0)}(p_i) \cup ... \cup obs^{(t)}(p_i)  
 = AP^{(0)}_{p_i} \cup ... \cup  AP^{(t)}_{p_i}
\end{array}}
$$

\noindent where $obs^{(t)}(p_i)$ represents the set of atomic propositions whose truth values
are known to process $p$ up to time step $t$.
Note that the above formula   does not take into consideration
the fact that processes communicate with each other
and that the observation power of processes
can be enhanced through communication. 
It is thus not limited to their local atomic propositions.
However, computing the set $obs^{(t)}(p_i)$ while considering remote observations
made by $p_i$ is crucial in our decentralized 
monitoring framework as it makes the inference rules 
\textbf{R3}-\textbf{R12} more powerful in the sense that
more information may be extracted when applying these rules.

Since processes use static  communication scheme in which the order at which observations
of processes are propagated is fixed between states,
they can then compute precisely the observation power
of each other.
Let $|\pi_{i, j}|$ be the number of communication rounds needed for process $p_i$ to influence the information state of process $p_j$,
where $|\pi_{i, j}|$ represents the length of the communication path between nodes $(p_i, p_j)$.
Note that in our framework we assume that each process can send at most
 one message at any communication round.
Therefore, when process $p_i$ makes a new observation $O_i$ at step $t$,
then process $p_j$ knows about $O_i$ at time step $(t + |p_{i, j}|)$.
So given a static communication scheme in which 
the length of communication path $|\pi_{i, j}|$
for any pair $(p_i, p_j)$ is  prior knowledge,
the observation power of a process $p_i$ at time step $t \geq 0$ 
can then be computed as follows:

\begin{equation} \label{ObsStatic}
 \begin{array}[t]{l}
obs^{(t)}(p_i) = ((AP_{p_i}^{(0)} \cup ... \cup AP_{p_i}^{(t)})~ \cup \\
 ~~~~~~~~~~(\forall_{ \tau = 0 ... t, j = 0..n \land j \neq i} ( \bigcup_{(t - \tau) \geq |\pi_{j, i}|} (AP_{p_j}^{(\tau)}))))
\end{array}
\end{equation}

That is, at time step $t$, process $p_i$ knows the truth values of its local 
propositions from time 0 to time $t$ and the truth values of all propositions
of process $p_j$ from time $0$ to time $\tau$, given that $(t - \tau) \geq |\pi_{j, i}|$. To demonstrate formula (\ref{ObsStatic}) let us consider the following example.

\begin{example} \label{Ex4}

Suppose that processes $\{A, B, C\}$ communicate with each other
using the fixed communication scheme $ comm = (A \rightarrow B \rightarrow C \rightarrow A)$
and that $comm$ is common knowledge among processes, 
where the direction of the arrow represents the direction of communication.  
We can then  compute for instance the observation power of processes $A, B$ and $C$ 
at step $t = 2$ as follows
$$
obs^{(2)}(A) = \{AP^{(0)}_{A},  AP^{(1)}_{A}, AP^{(2)}_{A}, AP^{(0)}_{C}, AP^{(1)}_{C}, AP^{(0)}_{B} \}
$$
$$
obs^{(2)}(B) = \{AP^{(0)}_{A},  AP^{(1)}_{A}, AP^{(0)}_{B}, AP^{(1)}_{B}, AP^{(2)}_{B}, AP^{(0)}_{C} \}
$$
$$
obs^{(2)}(C) = \{AP^{(0)}_{B},  AP^{(1)}_{B}, AP^{(0)}_{C}, AP^{(1)}_{C}, AP^{(2)}_{C}, AP^{(0)}_{A} \}
$$

\end{example}

Note that when the set $obs^{(2)}(B)$ contains the set $AP^{(0)}_{A}$
it means that $B$ knows the truth values of all propositions in the set 
$AP_{A}$ at time step 0. 
To demonstrate how inference rules
can be used in decentralized monitoring, let us consider the following example.


\begin{example}

Suppose that processes $\{A, B, C\}$ communicate with each other using the static communication
scheme $comm = (A \rightarrow B \rightarrow C \rightarrow A)$ and that 
$AP_A = \{a_1, a_2 \}$, $AP_B = \{b\}$, and $AP_C = \{c\}$.
The property to be monitored by processes is 
$\varphi = F (a_1 \land \neg a_2 \land b \land \neg c)$.
Let us denote the formula $(a_1 \land \neg a_2 \land b \land \neg c)$ by $\phi$.
We first construct a tableau for $\varphi$ using the tableau system 
of Section \ref{sec:tableau} (see Fig. \ref{fig: Ex}).
Suppose that at state $g_0$  (the initial state)
the propositions of the formula have the following truth values
 $a_1 = \top$, $a_2 = \bot$, $b = \top$ and $c = \bot$.
For brevity, we describe only the information propagated from process $A$ to process $B$ concerning state $g_0$.


\begin{figure}[h]
    \begin{center}
    \begin{tikzcd}  
    & F\phi \arrow{d} & \\
    & \phi\lor XF \phi \arrow{ld} \arrow{rd} &\\
    \phi \arrow{d}& &  F \phi \arrow{d} \\
    a_1, \neg a_2, b, c &   & \phi\lor XF  \\
    \surd &   & \surd
\end{tikzcd}
\end{center}
\caption{A tableau for $F \phi$, where $ \phi = (a_1 \land \neg a_2 \land b \land c)$} \label{fig: Ex}
\end{figure}

\begin{itemize}

\item At time step $t = 0$ process $A$ sends $(\phi = ?)$ to $B$.  Then using rule \textbf{R3}
process $B$ deduces that $a_1 = \top$ and $a_2 = \bot$ at state $g_0$ as $A$ observes both $a_1, a_2$.

\item At time step $t = 1$ process $A$ sends ($\phi= ?)$ to $B$. Then using rule \textbf{R3}
again process $B$ deduces that $a_1 = \top$ and $a_2 = \bot$ and $c = \top$ at state $g_0$ as $A$ observes $a_1, a_2$ and has already received a message from $C$ at time step $t=0$ about the truth value of proposition $c$ at state $g_0$. At this step, process $B$ knows truth values of all propositions at state $g_0$ as $b$ is locally observed by $B$.

\end{itemize}

\end{example}

\section{Propagating Observations in an Optimal Way} \label{sec:TableaRules}

In this section we describe an
algorithm that allows processes to propagate
their observations in an optimal way.
Since we assume that processes use a static communication scheme
based on some round-robin policy,
then they can compute precisely the observation power of each other.
Suppose that we use an RR communication policy in which
process $p$ sends its observations to process $q$. 
Then instead of allowing $p$ to propagate its entire knowledge
to $q$, it propagates only the set of observations
that are not known to $q$. We mean by observations the set of atomic propositions
whose definite truth values are known to $p$. 
The set of observations that $p$ needs to propagate to $q$ 
at time step $t$ can be computed as follow
$$
 \begin{array}[t]{l}
obs^{(t)} (p \rightarrow q) = (obs^{(t)}(p) - obs^{(t)}(q))  \\
 \hspace*{60 pt} = (AP^{(\tau)}_{k_p} - AP^{(\tau)}_{k_q}) \cup ... \cup (AP^{(t)}_{k_p} - AP^{(t)}_{k_q}) 
 \end{array}
$$
where the set $obs^{(t)}(p \rightarrow q)$ represents the set of new
 observations that process $p$ needs to propagate to process $q$, 
$AP^{(\tau)}_{k_p}$ represents the set of atomic propositions
whose definite truth values are known to $p$ at time step $\tau$,
where $\tau$ represents the earliest time step of the trace
at which a new observation has been made by process $p$. 
Note that process $p$ knows the set of propositions whose definite truth
values are known to $q$ from time 0 to time $t$.
This is due to the assumption that processes use a static 
communication scheme in
which the order at which observations are propagated is fixed between states.
It is interesting to note that process $p$ propagates not just
truth values of its locally observed propositions
but also truth values of propositions observed
by other processes. Let  $MinList$
be the set of minimal sets of  formulas whose truth values 
need to be propagated from steps $\tau$ to $t$.
The set  $MinList$ can be computed as follows
$$
MinList = MinList_{(\tau)} \cup .... \cup MinList_{(t)}
$$

We now summarize the steps that process $p$ 
needs to follow in order to compute the set $MinList$.

\begin{enumerate}

\item For each $k \in [\tau, t] $ process $p$ computes
the set $M_k = (AP^{(k)}_{k_p} - AP^{(k)}_{k_q})$: the set of propositions
at step $k$ whose truth values are known to $p$ but not known to $q$.

\item For each $M_k$ process $p$ computes the set $Minlist_{(k)}$ using Algorithm \ref{alg: MinimalSet}.

\item It then  combines the sets $Minlist_{(k)}$ for all $k \in [\tau, t] $ to obtain $MinList$.

\end{enumerate}




In our setting we assume that each process maintains a truth table
at each state of the monitored trace, which consists
of the set of formulas in the resulting tableau
of the monitored formula. 
The truth tables maintain 
only the set of non-atomic formulas that match the syntactic structure
of formulas in the inference rules  \textbf{R1-R12}, 
in addition to the set of atomic propositions of the main formula.
At each time step $\tau$, process $p$ examines the non-atomic
formulas in its truth tables to compute the minimal set
of formulas whose truth values need to be propagated (see Algorithm \ref{alg: MinimalSet}).
The truth table consists of three columns: 
the set of formulas, their truth values, and their unique index values.
We view the table $T_{\tau}$ as a set of entries of the form $(f, val, index)$. 
The set $CompoundFormulas$ used in the algorithm
represents the set of compound formulas
in the inference rules \textbf{R1-R12}.
The operation $DeductionSet (\phi)$ 
returns the set of atomic propositions in the formula $\phi$
whose definite truth values can be deduced.
As one can see Algorithm \ref{alg: MinimalSet} consists of two phases:
(a) the exploration phase
in which non-atomic formulas in the truth table $T_{\tau}$ are examined, 
and (b) the refinement phase in which redundant formulas 
in the list $MinList_{\tau}$ (if any) are removed.
Note that it is possible to have a formula whose
deduction list is subset of the 
deduction list of another formula or they have some atomic propositions in common.
The refinement phase is then used to detect such cases and remove redundant information.
This ensures that observations are propagated in an optimal way.

\begin{algorithm*} [h]
\begin{algorithmic}[1]
\State \textbf{Inputs} : $(M_{\tau}, T_{\tau}, CompoundFormulas)$
\State \textbf{Output} : $MinList_{\tau} = \emptyset$ 
\ForEach {$entry \in T_{\tau}$} \Comment{Exploration phase}
\If {$M_{\tau} \cap Atoms (entry.f) \neq \emptyset$}
\If {$MatchSynt (entry.f, \phi)$ for any $\phi \in CompoundFormulas$}
\If {$ entry.val =  \phi.val$}
\If {$DeductionSet (entry.f) \not \in DeductionSet (\phi)$  for any $\phi \in MinList_{\tau}$}
\State \textbf{add} $(entry.index, entry.val)$ \textbf{to} $ MinList_{\tau}$
\EndIf
\EndIf
\EndIf
\EndIf
\EndFor
\ForEach  {$\phi \in MinList_{\tau}$} \Comment{Refinement phase}
\If {$DeductionSet (\phi) \subseteq DeductionSet (\psi)$ for any $\psi \in MinList$}
\State \textbf{remove} $\phi$ \textbf{from} $MinList$ 
\EndIf 
\If{$DeductionSet (\psi) \subseteq DeductionSet (\phi)$ for any $\psi \in MinList$}
\State \textbf{remove} $\psi$ \textbf{from} $MinList$ 
\EndIf
\If {$DeductionSet (\phi) \subseteq (\bigcup_{\psi \in MinList} (DeductionSet (\psi)))$}
\State \textbf{remove} $\phi$ \textbf{from} $MinList$ 
\EndIf
\EndFor
\If {there exist $a \in M_{\tau}$ \textbf{such that} $ a \not \subseteq (\bigcup_{\phi \in MinList} (DeductionSet (\phi)))$}   
\State \textbf{add} $(index (a), val(a))$ to $MinList$ 
\EndIf
\State \textbf{return}  $MinList_{\tau}$ 
\end{algorithmic}
\caption{Computing minimal set of formulas for process $p$ at step $\tau$}  \label{alg: MinimalSet}
\label{alg:computingMinSet}
\end{algorithm*}


\section{Static Monitoring Approaches} \label{sec: staticApproach}

The knowledge state of processes in decentralized monitoring increases monotonically
over time due to local and remote observations.
It is therefore necessary to have an efficient
communication strategy that allows processes to
propagate only necessary observations. 
The propagation of monitoring
information from one peer to the other in our framework 
follows a static communication strategy based 
on a round-robin scheduling policy. 
The advantage of using a static
communication strategy in decentralized
monitoring is that processes can compute
precisely the knowledge state of each other
and communicate only new observations
to their neighbor processes (i.e. no redundant information will be propagated). 
This helps to reduce significantly the size 
of propagated messages, which is crucial when monitoring
large-scale distributed systems (systems with huge number of processes)
or systems running over wireless sensor networks.

To develop efficient static RR policies for decentralized monitoring
 we  use ranking functions to rank processes 
(i.e. assign a unique value to each process) using
 some interesting criteria that take into consideration
the observation power of processes and the syntactic
structure of the formula. The values 
assigned to the processes using ranking functions will be used then
 to specify the order of communication in
the round-robin policy. We consider
here two different ranking strategies.

\begin{enumerate}

\item For the first strategy, a process that contributes to the truth 
value of the formula via a larger number of propositions receives higher priority in the round-robin order; if multiple processes have the same number of propositions,
 then the order is fixed by an (externally provided) PID.   
The intuition behind choosing
such a strategy is that processes that 
observe more propositions of the formula
own more information about the global trace
of the system, and hence specifying the order of communication
in such a way may help to detect
any violation somewhat faster.


\item For the second strategy the tableau is considered, where we rank processes based on their observation power while taking into consideration the structure of the formula $\varphi$. 
Let $\mathcal{T}_{\varphi}$ be a tableau of $\varphi$. 
Then the process that contributes to a larger number of branches receives higher priority in the order of communication. 
Note that the tableau technique allows one to construct
a semantic tree for the monitored formula, where each branch of the tree
represents a way to satisfy the formula.
We believe that this strategy
would help to speed up the RV process.
\end{enumerate}

We now give an example to demonstrate how one can 
synthesize an RR policy using the above described 
strategies.

\begin{example}
Suppose we have a distributed system $S = \{A, B, C\}$ 
and a property $\varphi = G(a_1 \land \neg a_2 \land b_1) \lor F(b_2 \land c) $ 
 of $S$ that we would like to
 monitor in a decentralized fashion, 
 where  $AP_{A} = \{a_1, a_2\}$,
$AP_{B} = \{b_1, b_2\}$,
and $AP_{C} = \{c \}$.
From the syntactic structure of $\varphi$,
it is easy to see that the tableau $\mathcal{T}_{\varphi}$
constructed consists of three branches,
where one branch corresponds to formula $ G(a_1 \land \neg a_2 \land b_1)$ 
and two branches correspond to formula $F(b_2 \land c) $.
Hence the two strategies yield different RR policies as follows
$$
Comm_1 = A \rightarrow B \rightarrow C \rightarrow A
$$
$$
Comm_2 = B \rightarrow C \rightarrow A \rightarrow B
$$

\end{example}

\section{Decentralized Monitoring Algorithm}

\begin{table*} 
\begin{center}
\begin{tabular}{ |c|c|c|c|c|c|c|c|c|c|c|c|c| }
\hline
 &  \multicolumn{3}{|c|}{$|trace|$}& \multicolumn{3}{|c|}{$\# msg.$} & \multicolumn{3}{|c|}{$|msg.|$}  & \multicolumn{3}{|c|}{$|mem|$} \\
 \hline
$|\varphi|$ & BF & DM1 & DM2 & BF & DM1& DM2 & BF & DM1& DM2& BF & DM1& DM2  \\
\hline
1& 1.56 & 2.84& 2.64& 4.107 & 5.5 & 4.93& 86.5& 21.3 & 22.7 & 44.2 & 6.93 &  6.93  \\
\hline
2& 2.77 & 3.54 & 3.85& 6.285 & 6.9 &6.4 & 318 &  49.2 & 46.4 & 156 & 8.72 &  8.72\\
\hline
3& 6.79 & 9.36 & 8.56& 10.554 & 17.5 & 17.9 &3,540 & 166 & 173 &  458 & 12.4 & 12.4\\
\hline
4& 17.8 & 20.57 & 19.23& 17.667 & 37.6 & 36.9&  530 & 370& 357 & 1,100 & 13.3 &  13.3\\
\hline
5& 27.3 & 43.24 & 35.3& 28.125 & 74.7& 67.0& 4,650 & 877& 862 & 2630 & 14.4 &  14.4 \\
\hline
6& 61.2 & 64.16 &62.2 & 35.424 & 115.7& 113& 8,000 & 1,250 & 1,160 & 5,830 & 14.0 & 14.0\\
\hline
\end{tabular}
\end{center}
\caption{Benchmarks for 1000 randomly generated LTL formulas of size $|\varphi|$ (Averages)} \label{table: tableResult}
\end{table*}

\begin{table*} 
\begin{center}
\begin{tabular}{ |c|c|c|c|c|c|c|c|c|c|c|c|c| }
\hline
 &  \multicolumn{3}{|c|}{$|trace|$}& \multicolumn{3}{|c|}{$\# msg.$} & \multicolumn{3}{|c|}{$|msg.|$}  & \multicolumn{3}{|c|}{$|mem|$} \\
 \hline
$|\varphi|$ & BF & DM1 & DM2 & BF & DM1& DM2 & BF & DM1& DM2& BF & DM1& DM2  \\
\hline
abs &  4.65  & 5.26  & 5.15 & 4.46 & 6.25   & 6.15&  1,150 & 115   & 110  & 496 & 16.4 & 14.9 \\
\hline
exis &  27.9 &  32.4 & 31.5  &  19.7 & 22.5  & 20.8 &  1,100 & 525  & 521  & 376  & 22.6&21.8\\
\hline
bexis &  43.6 &  47.8 & 45.3 &  31.6 &  33.7 &  32.4 &  55,000 & 25784  & 25415  & 28,200& 27.5& 25.6\\
\hline
univ &   5.86&  6.8 & 6.2 &   5.92&  5.95 & 5.82 &   2,758 & 155 & 138  & 498& 24.7& 22.5\\
\hline
prec &   54.8 &  57.6 & 54.5 &   25.4& 27.9 & 26.9 &   8,625 & 785 & 755  & 663 & 35.4& 34.9\\
\hline
resp &   622 &  655 & 622 &   425&  545& 515 &    22,000 & 1225 & 1211 & 1,540 &19.8 & 17.5\\
\hline
precc & 4.11 &5.45 & 5.2 &  4.81& 6.25 &5.95 & 5,184 & 378&356 & 1,200 &17.4 & 15.7\\
\hline
respc & 427 & 452& 444&  381& 415 & 409& 9,000 &2875 & 2799&4,650 &23.4 & 22.1 \\
\hline
consc & 325 & 355& 324& 201& 243 & 234 & 7,200 & 1230& 1223 & 2,720 & 16.5& 15.8 \\
\hline
\end{tabular}
\end{center}
\caption{Benchmarks for 1000 generated LTL pattern formulas (Averages)} \label{table: tableResult2}
\end{table*}

Our decentralized monitoring algorithm consists of two phases: setup and monitor. The setup phase creates the monitors and defines their communication topology. The monitor phase allows the monitors to begin monitoring and propagating information to reach a verdict when possible. We first describe the steps of the setup phase.

\begin{itemize}

\item  Each process constructs a tableau $\mathcal{T}_{\varphi}$ for $\varphi$ 
using the method of Section \ref{sec:tableau}.

\item Each process then refines  the constructed tree
$\mathcal{T}_{\varphi}$ by removing redundant formulas and infeasible branches from the tree.
Note that in some cases the constructed tableau of an LTL formula may 
contain redundant formulas  due to repetitive loops (see Fig. \ref{Fig:EGP}).

\item Each process constructs a truth table that consists
of temporal formulas and compound formulas in $\mathcal{T}_{\varphi}$
that match the syntactic structure of the formulas in the inference rules \textbf{R1-R12},
in addition to the atomic formulas of  $\varphi$.

\item Each process assigns a unique index value
to each formula in the constructed truth table. We assume here that all
processes use the same enumerating procedure when assigning index values to the
formulas.

\end{itemize}

The advantage of assigning unique index values to the formulas in the constructed truth table  is that it allows processes to propagate truth values of formulas
 as pairs of the form $(idx(\phi), val)$, where $idx(\phi)$ 
 is the index value of the formula $\phi$ and   $val \in \mathbb{B}_3$,
 which helps to reduce the size of propagated messages.
We now summarize the actual monitoring steps in the form of an explicit algorithm that describes how local monitors operate and make decisions:

\begin{enumerate}

\item $[$Read next event$]$. Read next $\sigma_i \in \Sigma_i$ (initially each process reads $\sigma_0$).

\item $[$Compute minimal set of formulas to be transmitted$]$. Use Algorithm \ref{alg: MinimalSet} to derive $MinList_p^{g_i}$.

\item $[$Compute the receiving process$]$. For static approaches, the receiving
process of some process $p$ is fixed between states and computed according
to some round-robin communication policy, as described in Section \ref{sec: staticApproach}.

\item $[$Send truth value of formulas in $MinList_p^{g_i}$ $]$. Propagate the truth value of formulas in $MinList_p^{g_i}$  as pairs of the form $(idx(\phi), val)$ to the receiving process.


\item $[$Evaluate the formula $\varphi$ and return$]$.  If a definite verdict of $\varphi$ is found return it. That is, if $\varphi = \top$ return $\top$, if  $\varphi = \bot$ return $\bot$.

\item 	$[$Go to step 1$]$. If the trace has not been finished or a decision has not been made then go to step 1.

\end{enumerate}

We now turn to discuss the basic properties of our decentralized monitoring framework. 
Let $\models_{D}$ be the satisfaction relation on finite traces in the decentralized setting and $\models_{C}$ be the satisfaction relation on finite traces in the centralized setting, where both $\models_{D}$ and $\models_{C}$ yield values from the same truth domain. Note that in a centralized monitoring algorithm we assume that there is a central process that observes the entire global trace of the system being monitored, while in our  decentralized monitoring algorithm processes observe part of the trace, perform remote observation, and use some deduction rules in order to evaluate the property. The following theorems stating the soundness and completeness of our decentralized monitoring algorithm.

\begin{theorem} (\textbf{Soundness}). \label{soundness}
Let $\varphi \in LTL$ and $\alpha \in \Sigma^{*}$. Then $\alpha \models_{D} \varphi =\top/\bot \rimp \alpha \models_{C} \varphi =\top/\bot$.

\end{theorem}

Soundness means that all verdicts (truth values taken from a truth-domain) found by the decentralized monitoring algorithm for a global trace $\alpha$ with respect to the property $\varphi$ are actual verdicts that would be found by a centralized monitoring algorithm that have access to the trace $\alpha$. 

\begin{theorem} (\textbf{Completeness}).
Let $\varphi \in LTL$ and $\alpha \in \Sigma^{*}$. Then $\alpha \models_{C} \varphi =\top/\bot \rimp \alpha \models_{D} \varphi =\top/\bot$.

\end{theorem}

Completeness means that all verdicts found by the centralized monitoring algorithm 
for some trace $\alpha$ with respect to the property $\varphi$ 
will eventually be  found by the decentralized monitoring algorithm. 
The soundness and completeness of our monitoring strategies can be inferred from the 
soundness of inference rules \textbf{R1-R12},
Algorithm \ref{alg: MinimalSet}, and the tableau technique. 
The consistency property  (i.e.,  no two different monitors $M_i, M_j$ decide differently) of both approaches is inferred from the tableau approach.

\subsubsection*{Monitoring formulas with nested operators}

our framework can also handle formulas with nested operators 
(e.g., $G(\phi \rimp F \psi)$),
where the operators are expanded repeatedly 
using usual tableau expansion rules and the new LOOP checking rule and BRUNE rules 
introduced by Reynolds which can be used to halt the expansion
of infinite branches. The process stops
at trivial cases or when a certain loop condition is met 
(i.e. a fixed point has been reached). 
Processes can then exchange truth values of
atomic formulas, compound formulas,
or temporal formulas depending on the
tableau of the main formula.
We refer the reader to \cite{Reynolds2016} for
more details and examples.

\section{Implementation and Experimental Results}

We have evaluated our static monitoring strategies
against the LTL decentralized monitoring 
approach of Bauer and Falcone \cite{BauerF12},
in which the authors developed a monitoring algorithm for LTL based 
on the formula-progression technique \cite{Bacchus1996}.
The formula progression technique takes a temporal formula $\phi$ and a current assignment $I$ over the literals of $\phi$ as inputs and returns a new formula after acting  $I$ on $\phi$. The idea is to rewrite a temporal formula when an event $e$ is observed or received to a formula which represents the new requirement that the monitored system should fulfill for the remaining part of the trace.
To compare our monitoring approach
 with their decentralized algorithm,
we use the tableau system of Section \ref{sec:tableau},
which allows one to efficiently construct a semantic graph (tableau)
 for the input formula. 
We also use the tool DECENTMON3 \texttt{(http://decentmon3.forge.imag.fr/)} in our evaluation,
which is a tool dedicated to decentralized monitoring.
The tool takes as input multiple traces,
 corresponding to the behavior of a distributed system, and an LTL formula.
 The main reason behind choosing DECENTMON3 in our evaluation is that 
 it makes similar assumptions to our presented approach.
 Furthermore, DecentMon3 improves the original DecentMon tool developed in \cite{BauerF12}
  by limiting the growth of the size of local obligations and hence it may reduce the   size of propagated messages in decentralized monitoring. 
  We believe that by choosing the tool DECENTMON3 as baseline for comparison 
  we make the evaluation much fairer.

We denote by BF the monitoring approach of Bauer and Falcone,
 DM1 the first RR strategy of this paper in which processes are ordered
 according to the number of propositions they contribute to in the formula,
 and DM2 the second RR strategy of this paper in which processes are ordered
 according to the number of branches they contribute to in the tableau of the monitored formula. 
We compare the approaches against two benchmarks of formulas: 
randomly generated formulas (see Table \ref{table: tableResult}) and benchmark for patterns of formulas \cite{patternSite} (see Table \ref{table: tableResult2}). 
In Tables \ref{table: tableResult} and \ref{table: tableResult2}, the following metrics are used: $\# msg$, the total number of exchanged messages; $|msg|$, the total size of exchanged messages (in bits);  $|trace|$,  the average length of the traces needed to reach a verdict; and $|mem|$, the memory in bits needed for
the structures (i.e., formulas plus state for our algorithm).
 For example, the first line in Table \ref{table: tableResult} 
 says that we monitored 1,000 randomly generated LTL formulas of
size 1. On average, traces were of length 1.56 when  one 
of the local monitors in approach BF came to a verdict, 
and of length 2.84 and 2.64 when one of the monitors in DM1 and DM2  came to a verdict.

\subsection{Evaluation of randomly generated formulas}

Following the evaluation scheme of Falcone et al. \cite{BauerF12,FalconeCF14},
we evaluate the performance of each approach against a set of random
LTL formulas of various sizes.
For each size of formula (from 1 to 6), DECENTMON3
randomly generated 1,000 formulas. 
The result of comparing the three monitoring approaches can be seen in Table \ref{table: tableResult}. The first column of these tables shows the size of the monitored
LTL formulas. Note that we measure the formula size in terms of operator entailment inside it; for instance, $G (a \land b) \lor G (c \land d) \lor F (e)$ is of size 3. 
However, experiments show that operator entailment 
is more representative of how difficult it is to
progress it in a decentralized manner \cite{BauerF12,FalconeCF14}.
As shown in Table \ref{table: tableResult} 
our decentralized monitoring approaches 
lead to significant reduction on both
the size of propagated messages and the memory consumption
compared to the BF approach (i.e. the formula progression technique).
This demonstrates the effectiveness of the presented static round-robin approaches
and the inference LTL engine in decentralized monitoring.

\subsection{Benchmarks for Patterns of formulas}

We also compared the three approaches with more realistic
specifications obtained from specification patterns \cite{Dwyer1999}.
Table \ref{table: tableResult2} reports the verification results
for different kinds of patterns 
(absence, existence, bounded existence, universal, precedence,
response, precedence chain, response chain, constrained chain).
The actual specification formulas are available at \cite{patternSite}.
We generated also 1000 formulas monitored over
the same setting (processes are synchronous and reliable).
For this benchmark we generated formulas
as follows. For each pattern, we
randomly select one of its associated formulas.
Such a formula is ``parametrized''
by some atomic propositions from the alphabet of the distributed system
which are randomly instantiated.
For this benchmark (see Table \ref{table: tableResult2}), 
the presented approaches  lead also to significant reduction
on both the size of messages and the amount of memory consumption
compared to the optimized version of BF algorithm (DECENTMON3). 

\subsection{Discussion and Conclusions Based on Evaluation Results}

Comparing the decentralized monitoring
algorithms, the number of messages when using BF is always lower but the size of messages and the memory consumption is much bigger and  by several
orders of magnitude than our approaches. 
However, the approach DM2 showed better performance (on most of the cases) 
than DM1 in terms of the number and size of propagated messages.
This demonstrates the advantage of using tableau
in synthesizing an RR communication policy for decentralized monitoring.
A key drawback of using progression in decentralized monitoring (BF) is the
continuous growth of the size of local obligations with the length of the trace,
which imposes heavy overhead after a certain number of events. While progression
minimizes communication in terms of number of messages, it has the risk of 
saturating the communication device as processes 
send their obligations as rewritten temporal formulas.
On the other hand, processes in our  static monitoring approaches
(DM1 and DM2) propagate their observations as pairs of the form  $(idx(\phi), val)$ 
(rather that rewritten LTL formulas),
where  observations may be propagated as truth values of temporal 
and compound formulas. Furthermore, in our
approaches, processes propagate only new observations to their 
neighbor processes, which helps to reduce significantly the size of propagated messages.

\section{Related Work}

Several monitoring algorithms have been developed for verifying distributed systems at runtime \cite{Sen2004,BauerF12,ColomboF14,FalconeCF14,Scheffel14,MostafaB15}. They make different assumptions on the system model and thus target different kinds of distributed systems and they handle different specification languages.

Sen et al. \cite{Sen2004} propose a monitoring framework for safety properties of distributed systems using the past-time linear temporal logic. However, the algorithm is unsound. The evaluation of some properties may be overlooked in their framework. This is because monitors gain knowledge about the state of the system by piggybacking  on the existing communication among processes. That is, if processes rarely communicate, then monitors exchange very little information, and hence, some violations of the properties may remain undetected.

Bauer and Falcone \cite{BauerF12} propose a decentralized framework for runtime monitoring of LTL. The framework is constructed from local monitors which can only observe the truth value of a predefined subset of propositional variables. The local monitors can communicate their observations in the form of a (rewritten) LTL formula towards its neighbors. 
Mostafa and Bonakdarpour \cite{MostafaB15} propose similar decentralized LTL monitoring framework, but truth value of propositional variables rather than rewritten formulas are shared.

The work of Falcone et al. \cite{FalconeCF14} proposes a general decentralized monitoring algorithm in which the input specification is given as a deterministic finite-state automaton rather than an LTL formula. Their algorithm takes advantage of the semantics of finite-word automata, and hence they avoid the monitorability issues induced by the infinite-words semantics of LTL. They show that their implementation outperforms the Bauer and Falcone decentralized LTL algorithm \cite{BauerF12} using several monitoring metrics.

 Colombo and Falcone \cite{ColomboF16} propose a new way of organizing monitors called choreography, where monitors are organized as a tree across the distributed system, and each child feeds intermediate results to its parent. The proposed approach tries to minimize the communication induced by the distributed nature of the system and focuses on how to automatically split an LTL formula according to the architecture of the system.  

El-Hokayem and Falcone \cite{Hokayem2017} propose a new framework for decentralized monitoring with new data structure for symbolic representation and manipulation of monitoring information in decentralized monitoring. In their framework, the formula is modeled as an automaton where transitions of the monitored automaton are labeled with Boolean expressions over atomic propositions of the system.

A closer work to our work in this paper is the one of Basin et al. \cite{basin2015}, in which the authors use the 3-valued logic (strong Kleene logic) and an AND-OR graph to verify the observed system behavior at runtime with respect to specifications written in the real-time logic MTL. However, in our work we use a different construction to decompose and analyze the formula where we use the tableau graph, and we study the optimality problem in distributed monitoring with respect to both the number and size of transmitted messages at each state of the trace being monitored.  Furthermore, in our framework we allow monitors to use deduction rules to infer truth values of propositions from the messages they receive, where processes can propagate both atomic and compound formulas depending on both local and remote observations.

Our framework differs from the previous proposed decentralized frameworks
in the literature in that it is mainly based on the tableau construction,
where a semantic tree is constructed for the monitored formula using the tableau technique.
The constructed semantic tree is used then to synthesize an efficient communication strategy for the distributed system and to find the minimal ways to decompose the formula. Furthermore, our framework uses an inference engine for LTL which operates with different sets of inferences rules. The inference engine allows
processes to propagate their observations at each state of the monitored trace in an optimal way, while ensuring that only new observations are propagated. 

\section{Conclusion and Future Work}

We have presented an efficient decentralized monitoring framework for LTL
formulas using the tableau technique. The framework 
 consists of two parts: (a) an
algorithm that allows processes to compute at each state of the input execution trace the minimal set of formulas whose truth values need to be propagated, and (b) an LTL
inference engine that allows processes to extract the maximal amount of information
from the received messages.
The propagation of monitoring
information from one peer to the other in our framework 
follows a static communication strategy based 
on a round-robin scheduling policy. 
The advantage of using a static
communication strategy in decentralized
monitoring is that processes can compute
precisely the knowledge state of each other
and hence communicate only new observations. 
This helps to reduce significantly the size 
of propagated messages which is highly desirable
 when monitoring large-scale distributed systems. 
In future work, we aim to improve the static  approaches
by organizing processes into groups and then assign to each group
 a unique sub-formula of the main formula based on the semantics of the constructed
tableau of the formula. This should help to reduce further
the size of propagated messages.

\bibliographystyle{IEEEtran}
\bibliography{IEEEabrv,references}

\begin{thebibliography}{10}
\providecommand{\url}[1]{#1}
\csname url@samestyle\endcsname
\providecommand{\newblock}{\relax}
\providecommand{\bibinfo}[2]{#2}
\providecommand{\BIBentrySTDinterwordspacing}{\spaceskip=0pt\relax}
\providecommand{\BIBentryALTinterwordstretchfactor}{4}
\providecommand{\BIBentryALTinterwordspacing}{\spaceskip=\fontdimen2\font plus
\BIBentryALTinterwordstretchfactor\fontdimen3\font minus
  \fontdimen4\font\relax}
\providecommand{\BIBforeignlanguage}[2]{{%
\expandafter\ifx\csname l@#1\endcsname\relax
\typeout{** WARNING: IEEEtran.bst: No hyphenation pattern has been}%
\typeout{** loaded for the language `#1'. Using the pattern for}%
\typeout{** the default language instead.}%
\else
\language=\csname l@#1\endcsname
\fi
#2}}
\providecommand{\BIBdecl}{\relax}
\BIBdecl

\bibitem{ColomboPS08}
C.~Colombo, G.~J. Pace, and G.~Schneider, ``Dynamic event-based runtime
  monitoring of real-time and contextual properties,'' in \emph{Formal Methods
  for Industrial Critical Systems, 13th International Workshop, {FMICS} 2008,
  L'Aquila, Italy}, 2008, pp. 135--149.

\bibitem{Pike2011}
L.~Pike, S.~Niller, and N.~Wegmann, ``Runtime verification for ultra-critical
  systems,'' in \emph{Proceedings of the Second International Conference on
  Runtime Verification}, ser. RV'11, 2012, pp. 310--324.

\bibitem{DAngeloSSRFSMM05}
B.~D'Angelo, S.~Sankaranarayanan, C.~S{\'{a}}nchez, W.~Robinson, B.~Finkbeiner,
  H.~B. Sipma, S.~Mehrotra, and Z.~Manna, ``{LOLA:} runtime monitoring of
  synchronous systems,'' in \emph{12th International Symposium on Temporal
  Representation and Reasoning {(TIME} 2005), 23-25 June 2005, Burlington,
  Vermont, {USA}}, 2005, pp. 166--174.

\bibitem{Reynolds2016}
M.~Reynolds, ``{A New Rule for LTL Tableaux},'' in \emph{Symposium on Games,
  Automata, Logics and Formal Verification, GandALF 2016}, 2016, pp.
  287–--301.

\bibitem{Pnueli1977}
A.~Pnueli, ``The temporal logic of programs,'' in \emph{Proceedings of the 18th
  Annual Symposium on Foundations of Computer Science}, ser. SFCS '77.\hskip
  1em plus 0.5em minus 0.4em\relax IEEE Computer Society, 1977, pp. 46--57.

\bibitem{Bauer2011}
A.~Bauer, M.~Leucker, and C.~Schallhart, ``Runtime verification for ltl and
  tltl,'' \emph{ACM Transactions on Software Engineering and Methodology
  (TOSEM)}, pp. 14:1--14:64, 2011.

\bibitem{BauerF12}
A.~K. Bauer and Y.~Falcone, ``Decentralised {LTL} monitoring,'' in \emph{{FM}
  2012: Formal Methods - 18th International Symposium, Paris, France}, 2012,
  pp. 85--100.

\bibitem{ColomboF14}
C.~Colombo and Y.~Falcone, ``Organising {LTL} monitors over distributed systems
  with a global clock,'' in \emph{Runtime Verification - 5th International
  Conference, {RV} 2014}, 2014, pp. 140--155.

\bibitem{Pop06timinganalysis}
T.~Pop, P.~Pop, P.~Eles, Z.~Peng, and A.~Andrei, ``Timing analysis of the
  flexray communication protocol,'' in \emph{In Proc. of the ECRTS ’06},
  2006, pp. 203--216.

\bibitem{Gunzert1999}
M.~Gunzert and A.~N\"{a}gele, ``Component-based development and verification of
  safety critical software for a brake-by-wire system with synchronous software
  components,'' in \emph{Proceedings of the International Symposium on Software
  Engineering for Parallel and Distributed Systems}, 1999, pp. 134--.

\bibitem{Oliver2014AnalysisOD}
R.~S. Oliver, S.~S. Craciunas, and G.~Stoger, ``Analysis of deterministic
  ethernet scheduling for the industrial internet of things,'' \emph{2014 IEEE
  19th International Workshop on Computer Aided Modeling and Design of
  Communication Links and Networks (CAMAD)}, pp. 320--324, 2014.

\bibitem{Kleene29}
S.~C. Kleene, \emph{{Introduction to metamathematics}}, ser. Bibl.
  Matematica.\hskip 1em plus 0.5em minus 0.4em\relax Amsterdam: North-Holland,
  1952.

\bibitem{Bacchus1996}
F.~Bacchus and F.~Kabanza, ``Planning for temporally extended goals,'' in
  \emph{Proceedings of the Thirteenth National Conference on Artificial
  Intelligence}, 1996, pp. 1215--1222.

\bibitem{patternSite}
H.~Alavi, Avrunin, J.~G., Corbett, L.~Dillon, M.~Dwyer, and C.~Pasareanu,
  ``Specification patterns website,''
  \url{http://patterns.projects.cis.ksu.edu/}, 2011.

\bibitem{FalconeCF14}
Y.~Falcone, T.~Cornebize, and J.~Fernandez, ``Efficient and generalized
  decentralized monitoring of regular languages,'' in \emph{Formal Techniques
  for Distributed Objects, Components, and Systems}, 2014, pp. 66--83.

\bibitem{Dwyer1999}
M.~B. Dwyer, G.~S. Avrunin, and J.~C. Corbett, ``Patterns in property
  specifications for finite-state verification,'' in \emph{Proceedings of the
  21st International Conference on Software Engineering}, 1999, pp. 411--420.

\bibitem{Sen2004}
K.~Sen, A.~Vardhan, G.~Agha, and G.~Rosu, ``Efficient decentralized monitoring
  of safety in distributed systems,'' in \emph{Proceedings of the 26th
  International Conference on Software Engineering}, ser. ICSE '04.\hskip 1em
  plus 0.5em minus 0.4em\relax IEEE Computer Society, 2004, pp. 418--427.

\bibitem{Scheffel14}
T.~Scheffel and M.~Schmitz, ``Three-valued asynchronous distributed runtime
  verification,'' in \emph{International Conference on Formal Methods and
  Models for System Design (MEMOCODE)}, vol.~12.\hskip 1em plus 0.5em minus
  0.4em\relax IEEE, 2014.

\bibitem{MostafaB15}
M.~Mostafa and B.~Bonakdarpour, ``Decentralized runtime verification of {LTL}
  specifications in distributed systems,'' in \emph{2015 {IEEE} International
  Parallel and Distributed Processing Symposium}, 2015, pp. 494--503.

\bibitem{ColomboF16}
C.~Colombo and Y.~Falcone, ``Organising {LTL} monitors over distributed systems
  with a global clock,'' \emph{Formal Methods in System Design}, vol.~49, no.
  1-2, pp. 109--158, 2016.

\bibitem{Hokayem2017}
A.~El-Hokayem and Y.~Falcone, ``Monitoring decentralized specifications,'' in
  \emph{Proceedings of the 26th ACM SIGSOFT International Symposium on Software
  Testing and Analysis (ISTA)}, 2017, pp. 125--135.

\bibitem{basin2015}
D.~Basin, F.~Klaedtke, and E.~Zalinescu, ``{Failure-aware Runtime Verification
  of Distributed Systems},'' in \emph{35th IARCS Annual Conference on
  Foundations of Software Technology and Theoretical Computer Science (FSTTCS
  2015)}, vol.~45, 2015, pp. 590--603.

\end{thebibliography}

\end{document}